\documentclass[prl,superscriptaddress,reprint,floatfix]{revtex4-2}
\usepackage{graphicx,dcolumn,CJK,xparse,physics,color,bm,ulem}
\usepackage[colorlinks=true,breaklinks=true,allcolors=blue]{hyperref}
\usepackage{orcidlink}

\begin{document}
\title{Dissociation of bulk and entanglement phase transitions in the Haldane phase}
\begin{CJK*}{UTF8}{bsmi}
\author{Yu-Chin Tzeng (曾郁欽)\orcidlink{0000-0002-0380-1431}}
\email{Y.Tzeng-208@kent.ac.uk}
\affiliation{Physics of Quantum \& Materials Group, School of Engineering, Mathematics and Physics, University of Kent, Canterbury CT2 7NH, United Kingdom}
\affiliation{Department of ElectroPhysics and Center for Theoretical and Computational Physics, National Yang Ming Chiao Tung University, Hsinchu 300093, Taiwan}
\author{Gunnar M\"oller \orcidlink{0000-0001-8986-0899}}
\email{G.Moller@kent.ac.uk}
\affiliation{Physics of Quantum \& Materials Group, School of Engineering, Mathematics and Physics, University of Kent, Canterbury CT2 7NH, United Kingdom}

\begin{abstract}
Quantum entanglement provides a sensitive probe of topological phases and strong correlations in quantum many-body systems. We revisit the momentum-resolved entanglement spectrum (ES) of the spin-$\frac12$ XXZ ladder in the Haldane phase, whose SU(2)-symmetric spectrum has for over fifteen years been interpreted as a single des Cloizeaux--Pearson mode. Using momentum-resolved entanglement spectroscopy based on exact diagonalization of ladders with up to 40 spins, we instead resolve two distinct modes crossing at $k=\pi/2$ and forming a sharp cusp rather than a single $\sin|k|$ branch, revising the SU(2)-symmetric ES. We then introduce explicit SU(2) symmetry breaking through an XXZ anisotropy. The ES undergoes a phase transition at the isotropic point, while the physical ladder remains in the Haldane phase until a distinct bulk transition at larger anisotropy, establishing a dissociation between bulk and entanglement phase transitions. Between the entanglement and bulk critical points, no gapless entanglement branch remains that can be interpreted as a Haldane-edge excitation, signalling breakdown of the Li--Haldane correspondence. In the easy-plane regime, the ES develops an Anderson tower of states, and the entanglement ground state exhibits long-ranged spin correlations on the finite chains studied, consistent with emergent U(1)-broken order and incompatible with conventional short-range one-dimensional Hamiltonians satisfying the assumptions underlying the Lieb--Schultz--Mattis and Mermin--Wagner--Hohenberg--Coleman theorems. Together, these results establish entanglement-only quantum criticality as a distinct manifestation of critical behaviour encoded in the entanglement description.
\end{abstract}

\maketitle
\end{CJK*}

\textit{Introduction}.--
Quantum entanglement has become a key probe of quantum phases beyond symmetry breaking~\cite{Amico:RMP2008, Laflorencie:review2016, Pollmann2010, Foss-Feig_PRL2022}, including in non-Hermitian systems~\cite{Poyao2020,Tu2022,symm-res,Hsieh2023,Ju2024}. 
In particular, the entanglement spectrum (ES), defined as the set of eigenvalues $\{\xi_i\}$ of the entanglement Hamiltonian $H_E$~\cite{Li-Haldane,Vidal2015, EH:review2022, tomography,Quantum, BW1975,BW1976,Giudici2018,Rottoli2025, Pollmann_PRX2025} through the relation 
\begin{equation}
\rho_A=\frac{e^{-\beta H_E}}{Z},
\end{equation}
has proven especially valuable in revealing topological order and edge physics~\cite{Li-Haldane,Vidal2015}. Here, $\rho_A=\mathrm{Tr}_B|\psi_0\rangle\langle\psi_0|$ is the reduced density matrix of subsystem $A$, obtained by tracing out subsystem $B$ from the ground state $|\psi_0\rangle$~\footnote{The inverse temperature $\beta=1$ is set to unity, and $Z=\mathrm{Tr}(e^{-\beta H_E})$ is a normalization constant ensuring $\mathrm{Tr}(\rho_A)=1$. With $\xi_i=-\ln\omega_i-\ln Z$, only differences $\xi_i-\xi_0$ are meaningful, as the additive constant $\ln Z$ is unknown and extensive. Literature ``$\xi_0$'' usually means $-\ln\omega_0$, where $\omega_0$ is the largest eigenvalue of $\rho_A$. Assuming $(\ln Z)/L=\mbox{const.}+O(1/L)$, finite-size scaling of ``$\xi_0/L$'' as an energy density can include $O(1/L)$ and higher-order corrections.}.
According to the Li--Haldane conjecture~\cite{Li-Haldane}, the low-lying ES reflects the conformal field theory (CFT)-like excitation spectrum of edges in topological phases, establishing what is now referred to as the edge-ES correspondence~\cite{Vidal2015}.

The two-leg ladders provide a natural setting to extend the Li--Haldane conjecture to extensive bipartitions~\cite{Poilblanc2010,PEPS, Hsieh2014,Santos2016}, where the entanglement cut divides the system along its length rather than across its edge~\cite{Poilblanc2010,PEPS}. 
The two-leg ladder therefore provides a natural setting to test the extent to which the Li--Haldane conjecture continues to hold under extensive bipartitioning, and to investigate how the entanglement spectrum evolves across the XXZ phase diagram.
A seminal study by Poilblanc~\cite{Poilblanc2010} investigated the ES of the spin-1/2 Heisenberg ladder in the Haldane phase (up to $L=14$) and reported a smooth $\sin|k|$-like dispersion under momentum-resolved spectroscopy. The observed $\sin|k|$-like dispersion was seen as a signature of gapless edge modes, similar to the des Cloizeaux--Pearson (dCP) mode of a critical spin-1/2 chain~\cite{desCloizeaux,book:Mikeska2004}, and has influenced the standard interpretation of the ES in topological ladder systems.
On the other hand, a dCP-like entanglement dispersion also appears in the trivial rung-singlet phase~\cite{Poilblanc2010}, and was originally interpreted on equal footing with the Haldane phase, suggesting a unified edge interpretation across both phases.

\begin{figure*}[t]
\includegraphics[width=2in]{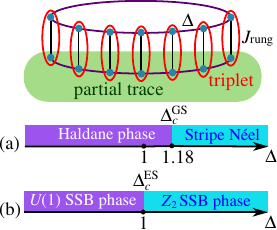}
\includegraphics[width=5in]{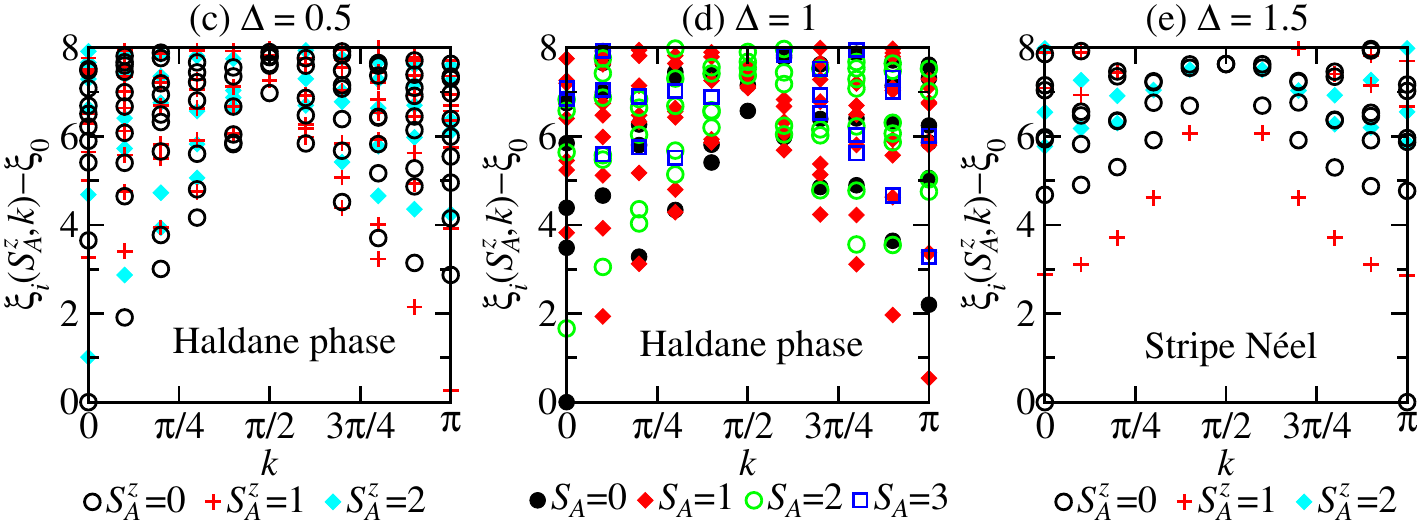} 
\caption{Summary of main results for a spin-1/2 ladder with $L=20$, $J_\mathrm{rung}=-6$, and $0.5\leq\Delta\leq1.5$. Only half the Brillouin zone ($0\leq k\leq\pi$) is shown because the spectrum is symmetric under $k\to-k$.
(a) Ground-state phase diagram of the full ladder, showing a quantum phase transition at $\Delta=\Delta_c^\mathrm{GS}\approx1.18$ between the Haldane and Stripe-N\'eel phases.  
(b) Phase diagram of the entanglement ground state, exhibiting a second-order quantum phase transition at $\Delta=\Delta_c^\mathrm{ES}\approx1$, identified via the entanglement fidelity susceptibility density $\chi_\mathrm{ES}$.  
(c) At $\Delta=0.5$, a gapped quadratic mode at $k=\pi$ in the $S_A^z=0$ sector demonstrates a violation of the LSM constraint for short-range systems. In addition, the ground-state degeneracy across $S_A^z$ sectors in the thermodynamic limit signals U(1) symmetry breaking, in conflict with the Mermin--Wagner--Hohenberg--Coleman theorem.  
(d) At the SU(2)-symmetric point $\Delta=1$, the $k=\pi$ branch becomes gapless and linear. The cusp at $k=\pi/2$ indicates a departure from the smooth $\sin|k|$ dispersion, revealing two distinct branches rather than a single dCP-like mode.
(e) At $\Delta=1.5$, the doubly degenerate ES ground state indicates spontaneous $Z_2$ symmetry breaking.
For spectra at intermediate $\Delta$ values, see Fig.\ref{fig:S2} in the End Matter.
}
\label{fig:1}
\end{figure*}

Our new results fundamentally challenge this interpretation by revealing that the entanglement Hamiltonian forms an autonomous many-body system with its own phase diagram, rather than a faithful representation of edge excitations. In the trivial phase, the $\sin|k|$ shape arises from an effective thermal mapping between the entanglement Hamiltonian and the subsystem Hamiltonian, $H_E\approx\beta_\mathrm{eff}H_A$, as analytically confirmed by perturbation theory~\cite{Lauchli_2012,Reconstruct} and extended to non-Hermitian cases~\cite{PeiYun_2024}. 
In contrast, we show that the ES in the Haldane phase does not follow a smooth dCP-like dispersion; instead, it consists of two branches whose gaps evolve with $\Delta$, signaling a quantum phase transition of the entanglement Hamiltonian at $\Delta_c^{\mathrm{ES}}\approx1$, independent of the bulk Haldane phase, as shown in Fig.~\ref{fig:1}(a-b). This distinction undermines the supposed universality of the dCP mode and calls for a revision of the 15-year-old established picture~\cite{Poilblanc2010}  
that has shaped the field's understanding of ladder entanglement spectra.


These observations highlight the need to reassess the structure and universality of the ES more broadly. Recent numerical advances have enabled momentum-resolved studies of spin ladders, but with important limitations: quantum Monte Carlo can access large system sizes in the rung-singlet phase~\cite{QMC2014,QMC2023,QMC:2D,QMC2025,QMC:IsingLadder}, although its accuracy in the Haldane phase is severely limited by the notorious sign problem; meanwhile, DMRG captures gapless spectra in SU(2)-symmetric ladders~\cite{extensive2024} but has mainly focused on the AKLT point and notably misses the $k=\pi/2$ mode. 
Crucially, the emergent non-locality we identify implies a violation of the entanglement area law~\cite{Kuwahara2020}, posing intrinsic challenges for such tensor-network-based approaches in the Haldane phase.

Extensions of Lieb--Schultz--Mattis (LSM) theorem~\cite{book:Auerbach,LSM1961, LSM:Oshikawa1997,LSM:Oshikawa2000,Kobayashi2019,Yao2021, LSM:Yao2024,Zang2024,Su2025, Ma2024, Liu2025} to open systems~\cite{Kawabata:LSM2024} and to entanglement Hamiltonians~\cite{NSR2025} have recently proposed that the ES may obey universal constraints, sparking renewed interest in such fundamental limits; 
however, as the present work reveals, long-standing assumptions about the ES structure can obscure how and when such LSM-type constraints genuinely apply to entanglement Hamiltonians. These challenges motivate an exact and momentum-resolved approach capable of disentangling thermal mappings from truly universal features.

In this Letter, we therefore study the ES of spin ladders using exact diagonalization (ED)~\cite{HQLin1990,Laflorencie2004}. This method provides full momentum resolution and unbiased access to the entanglement Hamiltonian. By introducing a continuous tuning parameter $\Delta$ for an XXZ deformation along the ladder legs, we identify three distinct regimes in the ES, as shown in Fig.~\ref{fig:1}(b): a U(1) symmetry-broken phase for $\Delta<1$, a critical point at $\Delta=1$, and a gapped $Z_2$ phase for $\Delta>1$. We then characterize the symmetry-broken phase and perform finite-size scaling, which reveals an entanglement transition clearly separated from the bulk transition at $\Delta_c^{\mathrm{GS}}\simeq 1.18$.
This separation demonstrates that the entanglement and bulk phase diagrams evolve differently under XXZ anisotropy, signalling a breakdown of the conventional Li--Haldane correspondence.

\textit{Model}.--
We study a spin-1/2 XXZ two-leg ladder defined by $H=H_A+H_B+H_{AB}$ with periodic boundary conditions,
\begin{align}
&H_A\!=\!\sum_{j=1}^L(S_{A,j}^x S_{A,j+1}^x \!+\! S_{A,j}^y S_{A,j+1}^y \!+\!\Delta S_{A,j}^z S_{A,j+1}^z),\label{eq:H_A}\\
&H_{AB}\!=\!J_{\mathrm{rung}}\sum_{j=1}^L\mathbf{S}_{A,j}\cdot\mathbf{S}_{B,j},\label{eq:H_AB}
\end{align} 
where $H_B$ is identical to $H_A$. With fixed $J_\mathrm{rung}=-6$, the rungs form effective spin-1's, and tuning $0.5\!\leq\!\Delta\!\leq\!1.5$ drives a quantum phase transition from the Haldane to the Stripe-N\'eel phase at $\Delta_c^\mathrm{GS}\!\approx\!1.18$~\cite{Hijii2005}. The global ground state always lies in the sector of total magnetization $(S_A^z+S_B^z)=0$ and total momentum $(k+q)\bmod2\pi=0$, where $S_A^z=\sum_{j=1}^LS_{A,j}^z$ and the momentum $k$(or $q$)$=2\pi j/L$ is defined within subsystem~A (B). Meanwhile, the entanglement Hamiltonian on a single leg exhibits even-odd momentum shifts. Following the convention in Ref.~\cite{Poilblanc2010}, we shift the momentum as $k\to k-\pi$ for system sizes $L=4m+2$. We denote entanglement levels by $\xi_i(S_A^z,k)$, with entanglement ground state energy $\xi_0=\xi_0(0,0)$. ED up to 40 spins is performed~\cite{HQLin1990,Laflorencie2004}, exploiting magnetization and translational symmetry, reaching Hilbert space dimension $6.89\times 10^9$.

\begin{figure}[t]
\centering
\includegraphics[width=\linewidth]{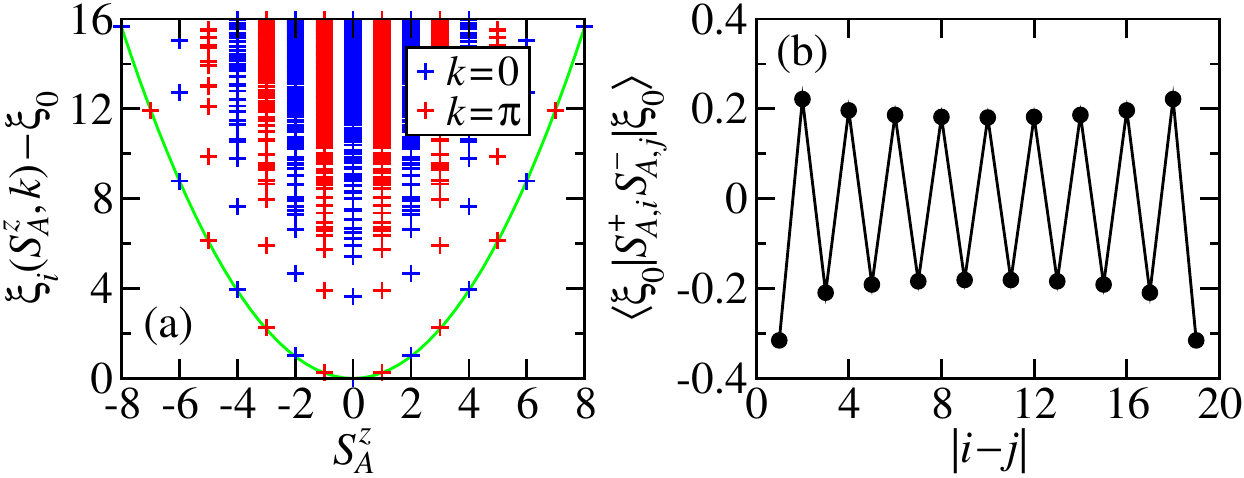}
\caption{
(a) Anderson tower of states in the ES for $\Delta=0.5$, demonstrating the characteristic structure of U(1) continuous symmetry breaking. The green line shows a parabolic fit to the lowest entanglement level, Eq.~\eqref{eq:tos} 
with $L=20$ and $\kappa\simeq0.1$.
(b) Correlation function in the entanglement ground state for $\Delta=0.5$ provides direct numerical evidence for the presence of long-range order. 
}\label{fig2:tos}
\end{figure}

\begin{figure}[t]
\centering
\includegraphics[width=\linewidth]{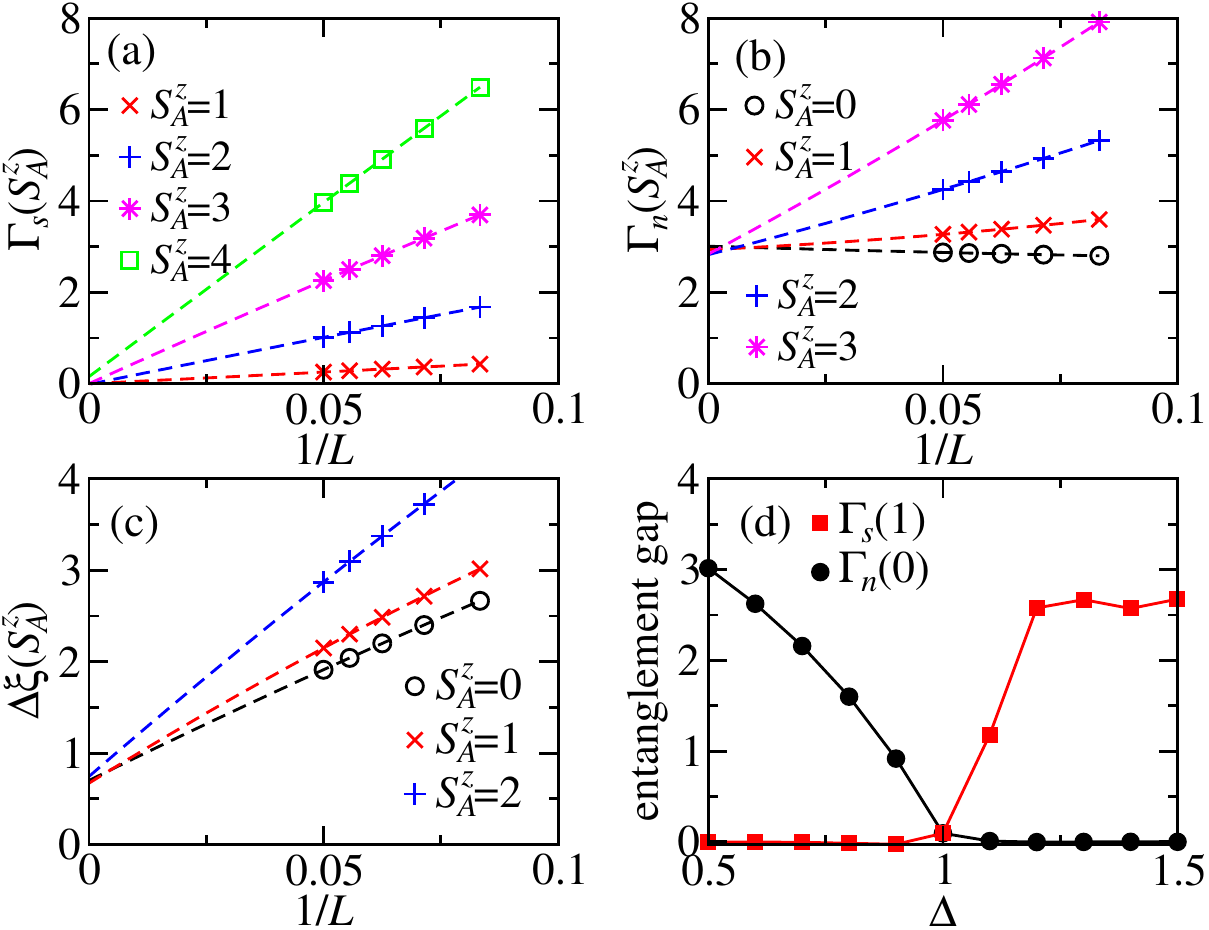}
\caption{
Finite-size scaling of entanglement gaps at $\Delta=0.5$. All the scaling use quadratic fitting functions.
(a) $\Gamma_s(S_A^z)$ scale as $1/L$ and vanish in the thermodynamic limit for all $S_A^z$, consistent with the Anderson tower in Fig.~\ref{fig2:tos}(a).  
(b) Quadratic-mode gaps $\Gamma_n(S_A^z)$ remain finite as $L\to\infty$, demonstrating a violation of the LSM theorem.
(c) Linear-mode gaps extrapolate to a finite offset, confirming that the entanglement ground state in the $S_A^z=0$ sector remains unique and non-degenerate.  
(d) Thermodynamic-limit values of the spin gap $\Gamma_s(1)$ and the neutral gap $\Gamma_n(0)$ versus $\Delta$, showing a crossing near $\Delta=1$.
}\label{fig3:gap}
\end{figure}

\textit{Results}.--
In Fig.~\ref{fig:1}(c-e) we present overviews of ES for three representative anisotropies $\Delta=0.5,1$, and $1.5$, respectively. For $\Delta=0.5$ [Fig.~\ref{fig:1}(c)], the ES exhibits interlaced linear and quadratic branches alternating between $k=0$ and $k=\pi$ across $S_A^z$ sectors, forming a quasi-degenerate tower. This structure signals emergent U(1) spontaneous symmetry breaking (SSB) in $H_E$, in apparent violation of the Mermin--Wagner--Hohenberg--Coleman theorem for systems with short-range interactions and Lorentz invariance~\cite{Mermin-Wagner,Hohenberg1967,Coleman1973,Nakanishi1977}. While it is now understood that continuous symmetry breaking can occur even in short-range 1D systems that lack Lorentz invariance~\cite{Watanabe2024}, the specific spectral features we observe point towards a different origin. The quadratic scaling of the Anderson tower, $\Gamma_s(S_A^z)\propto (S_A^z)^2/L$ [see Eq.~\eqref{eq:tos}], and the persistence of a finite neutral gap $\Gamma_n(0)$ [see Fig.~\ref{fig3:gap}(b,d)] are both characteristic hallmarks of systems with effective long-range interactions. This strongly suggests that the emergent non-locality of the entanglement Hamiltonian is the primary mechanism circumventing these fundamental constraints in our system. Within the $S_A^z=0$ sector, the gapped quadratic branch at $k=\pi$ further constitutes an apparent violation of the LSM-type constraint~\cite{NSR2025}. 
We emphasize that this is not a violation of the rigorous LSM theorem, but rather a demonstration that the entanglement Hamiltonian fails to satisfy the theorem's locality assumptions---as we elaborate in the End Matter.

At the SU(2)-symmetric point $\Delta=1$ [Fig.~\ref{fig:1}(d)], the ES is organized by total spin $S_A$, forming multiplets consistent with full spin-rotation symmetry. Two distinct linear branches at $k=0$ and $k=\pi$ intersect at a sharp cusp near $k=\pi/2$. Earlier studies~\cite{Poilblanc2010} interpreted this as a smooth $\sin|k|$ dispersion, but the cusp was missed due to limited resolution and the fact that $k=\pi/2$ appears only for subsystem lengths $L=4m$. Larger sizes reveal that the lowest level at $k=\pi/2$ is a singlet ($S_A=0$), with the triplet ($S_A=1$) just above, confirming that the spectrum consists of two crossing branches rather than a single dCP-like mode. This shows that, despite superficial similarities, the entanglement Hamiltonian is not truly analogous to the physical Hamiltonian of the edge.

For $\Delta=1.5$, deep in the easy-axis regime, both the $k=0$ and $k=\pi$ branches are quadratic and gapped, while the entanglement ground state becomes twofold degenerate between these two momenta, consistent with $Z_2$ symmetry breaking. In this case the LSM-type constraint is revived~\cite{NSR2025}, suggesting that the entanglement Hamiltonian effectively reduces to a short-range interaction. This contrasts sharply with the easy-plane regime at $\Delta=0.5$, where long-range interactions emerge.

As shown in Fig.~\ref{fig2:tos}(a), the ES at finite size $L=20$ and $\Delta=0.5$ develops an Anderson tower of states~\cite{Anderson1952}, with the lowest level in each $S_A^z$ sector following a well-defined quadratic dependence that extends to large $|S_A^z|$. 
\begin{equation}\label{eq:tos}
\Gamma_s(S_A^z)=\frac{(S_A^z)^2}{2\kappa L}
\end{equation}
In addition, Fig.~\ref{fig2:tos}(b) demonstrates that the entanglement ground state exhibits long-range correlations, directly evidencing long-range order.

\textit{Entanglement gap scaling}.--
We now perform a finite-size scaling analysis of low-lying entanglement gaps to clarify the excitation structure. We define three types of gaps: (i) the spin gap: $\Gamma_s(S_A^z)=\xi_0(S_A^z,k)-\xi_0$, with $k=0,\pi$ for even and odd $S_A^z$, respectively; (ii) the quadratic gap: $\Gamma_n(S_A^z)$, given by the lowest levels at $k=\pi,0$ for even and odd $S_A^z$, respectively. For $S_A^z=0$, $\Gamma_n(0)=\xi_0(0,\pi)-\xi_0$ defines the neutral gap. (iii) the linear-mode gap: $\Delta\xi(S_A^z)$, taken at $k=2\pi/L$ and $k=\pi-2\pi/L$ for even and odd $S_A^z$, respectively.
Figure~\ref{fig3:gap}(a) demonstrates that $\Gamma_s(S_A^z)$ exhibits $1/L$ scaling for all $S_A^z$, implying a macroscopically degenerate ground state as $L\to\infty$. This scaling behavior matches the Anderson tower spectrum identified in Fig.~\ref{fig2:tos}(a), perfectly fitting the functional form of Eq.~\eqref{eq:tos} with $\kappa\simeq0.1$.

Figure~\ref{fig3:gap}(b) shows that the quadratic gaps $\Gamma_n(S_A^z)$ remain finite as $L\to\infty$. In particular, for $S_A^z=0$, the LSM theorem states that in a translationally invariant system with U(1) spin-rotation symmetry and half-integer spins per unit cell, the ground state and an excitation within the same $S_A^z=0$ sector but with momentum shifted by $\pi$ must become degenerate in the thermodynamic limit, unless the Hamiltonian involves long-range interactions~\cite{book:Auerbach,LSM1961, LSM:Oshikawa1997,LSM:Oshikawa2000,Kobayashi2019, Yao2021,LSM:Yao2024,Zang2024,Su2025, Ma2024, Liu2025}.
Numerically, however, $\Gamma_n(0)$ remains finite for the parameters considered here, indicating a breakdown of the LSM scenario for the entanglement Hamiltonian~\cite{NSR2025} (see End Matter for details). This persistence of a finite gap shows that the entanglement Hamiltonian evades the short-range assumptions of the theorem. Such a breakdown is consistent with the emergence of effective long-range interactions, which are known to stabilize continuous symmetry breaking in 1D quantum systems~\cite{Maghrebi2017}.

Figure~\ref{fig3:gap}(c) further shows that the gaps of the linear mode extrapolate to a finite offset, confirming that the entanglement ground state in the $S_A^z=0$ sector remains unique and non-degenerate. This linear mode can be regarded as a Goldstone-like mode associated with U(1) symmetry breaking, but with a finite gap induced by long-range interactions.

Finally, Fig.~\ref{fig3:gap}(d) compares the thermodynamic-limit extrapolations of the spin gap $\Gamma_s(1)$ and the neutral gap $\Gamma_n(0)$ across a range of $\Delta$. In the easy-plane regime, the spin gap vanishes while the neutral gap remains finite, whereas in the easy-axis regime, the opposite holds. This crossing signals an entanglement quantum phase transition. Notably, the full ladder remains in the symmetry-protected topological (SPT) Haldane phase throughout this parameter range, since the XXZ anisotropy preserves the relevant protecting symmetries, i.e.~parity, time-reversal, and $Z_2\times Z_2$ symmetries~\cite{Pollmann2010}. Thus, the transition observed in the ES does not coincide with the bulk critical point, and in fact must be disconnected from it~\cite{Chandran2014}. To sharpen this distinction and precisely locate the entanglement critical point, we employ the fidelity susceptibility~\cite{You2007}, a powerful diagnostic that remains effective even without explicit knowledge of the underlying Hamiltonian, and has been broadly applied~\cite{fidelity_review}, including in recent studies of non-Hermitian systems~\cite{Tzeng2021,Tu2023}.

\begin{figure}[t]
\centering
\includegraphics[width=\linewidth]{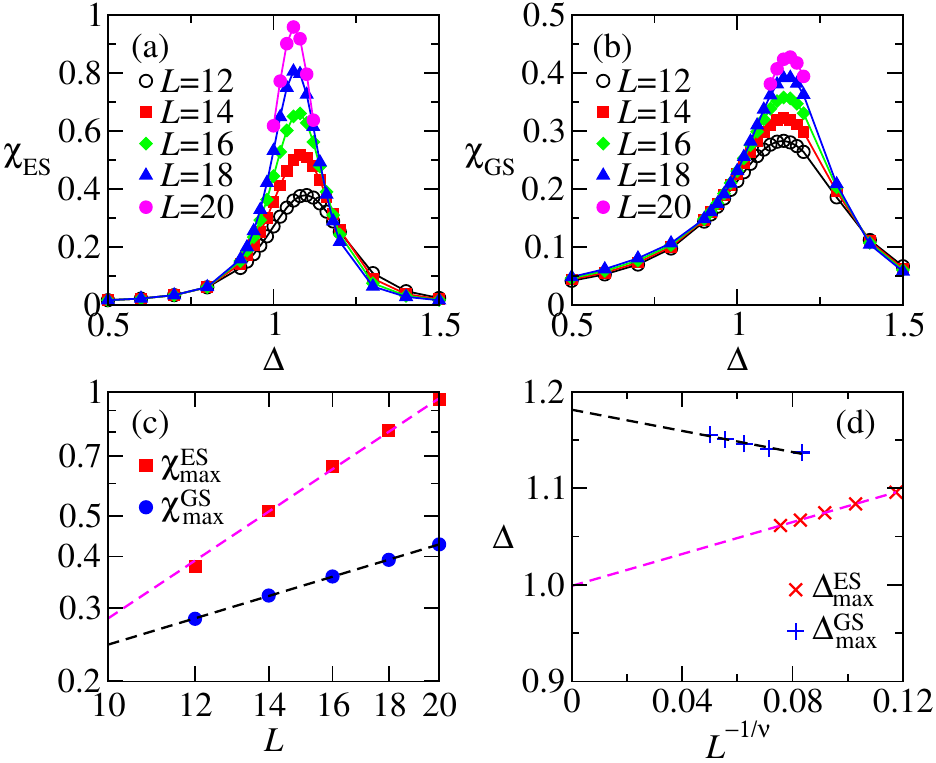} 
\caption{
(a) Entanglement fidelity susceptibility density $\chi_\mathrm{ES}$ exhibits a peak near the SU(2)-symmetric point $\Delta=1$.
(b) Ground-state fidelity susceptibility density $\chi_\mathrm{GS}$ peaks at the bulk Haldane-N\'eel transition.
(c) Doubly-logarithmic plot of peak values reveals that $\chi^\mathrm{ES}_\mathrm{max}$ and $\chi^\mathrm{GS}_\mathrm{max}$ scale as power laws with system size $L$, consistent with second-order phase transitions. The dashed lines are least-squares linear fits to the data.
(d) Extrapolation of the peak positions yields the entanglement critical point $\Delta_c^\mathrm{ES}=0.9988\pm0.013$ and the bulk critical point $\Delta_c^\mathrm{GS}=1.1813\pm0.058$, with the correlation length exponents $\nu\approx1.16$ and $\nu\approx1$, respectively.
}\label{fig:fsus}
\end{figure}

\textit{Fidelity susceptibility}.--
We define the entanglement fidelity susceptibility density as 
\begin{equation}
    \chi_\mathrm{ES}=\frac{1-|\langle\xi_0(\Delta)|\xi_0(\Delta+\delta)\rangle|^2}{\delta^2L}
\end{equation}
which quantifies how sensitively the entanglement ground state $|\xi_0\rangle$ responds to a small change $\delta=10^{-3}$ in the anisotropy parameter $\Delta$. For comparison, we also compute the ground-state fidelity susceptibility density of the full system, $\chi_\mathrm{GS}=[1-|\langle\psi_0(\Delta)|\psi_0(\Delta+\delta)\rangle|^2]/(\delta^2L)$. As shown in Fig.~\ref{fig:fsus}(a) and (b), $\chi_\mathrm{GS}$, peaks at $\Delta>1$, consistent with the known bulk quantum phase transition from the Haldane to the Stripe N\'eel phase~\cite{Hijii2005}. In contrast, $\chi_\mathrm{ES}$ displays a distinct peak near the SU(2)-symmetric point $\Delta=1$, signaling a separate transition in the entanglement Hamiltonian that occurs entirely within the Haldane phase.

Figure~\ref{fig:fsus}(c) shows that the peak values $\chi^\mathrm{GS}_\mathrm{max}$ and $\chi^\mathrm{ES}_\mathrm{max}$ both grow with system size $L$ following power-law scaling, a hallmark of conventional second-order phase transitions. 
This behavior clearly contrasts with the Berezinskii-Kosterlitz-Thouless (BKT) transitions realized in the ``parent system'' of a single 1D XXZ chain Eq.\eqref{eq:H_A}, where the fidelity susceptibility density typically grows with logarithmic corrections or remains nearly flat for modest sizes ($L\lesssim20$) and rarely exhibits sharp features in ED studies~\cite{fidelity_BKT,MFYang2007}.

To locate the critical points more precisely, we apply standard finite-size scaling analyses~\cite{Tzeng2008a} to the peak positions, as shown in Fig.~\ref{fig:fsus}(d). The extrapolated values identify two distinct transitions: $\Delta_c^\mathrm{ES}=0.9988\pm0.013$ for the entanglement Hamiltonian and $\Delta_c^\mathrm{GS}=1.1813\pm0.058$ for the bulk system. The absence of any anomaly in $\chi_\mathrm{ES}$ at the bulk transition point further emphasizes that the entanglement Hamiltonian can exhibit its own symmetry-driven phase transition, independent of the criticality of the underlying physical system.

The separation between $\Delta_c^\mathrm{ES}$ and $\Delta_c^\mathrm{GS}$ implies the existence of an intermediate regime, $\Delta_c^\mathrm{ES}<\Delta<\Delta_c^\mathrm{GS}$, in which the bulk system remains in the Haldane phase, yet the ES has already undergone a transition. In this regime, the entanglement Hamiltonian exhibits spontaneous $Z_2$ symmetry breaking, featuring a doubly degenerate ground state and gapped, quadratic modes at both $k=0$ and $k=\pi$, while simultaneously, no linear mode survives in the ES that could be interpreted as representing an edge excitation. This challenges the extension of the original Li--Haldane conjecture~\cite{Li-Haldane} to extensive bipartitions, which posits a correspondence between the ES and physical edge modes in topological phases. The absence of such correspondence in this regime suggests that the entanglement Hamiltonian captures physics beyond conventional boundary theories, and may not adhere to the same field-theoretic constraints.

\textit{Conclusion}.--
We have revisited the entanglement spectrum~(ES) of the spin-1/2 ladder in the Haldane phase, demonstrating that the long-standing picture of a single dCP-type $\sin|k|$ dispersion is an artifact of SU(2) symmetry and finite-size effects. 
Large-scale ED resolves two distinct branches at $k=0$ and $k=\pi$, with gap openings evolving as $\Delta$ is varied, signaling that the entanglement Hamiltonian undergoes a quantum phase transition at $\Delta_c^{\mathrm{ES}}\approx1$, well inside the bulk Haldane phase, and behaves as an autonomous many-body system.
This refinement of the spectrum exposes the limits of the Li--Haldane correspondence for extensive bipartitions: the entanglement Hamiltonian can enter a gapped $Z_2$ phase with no linear mode that could represent a physical edge excitation, even while the bulk remains a symmetry-protected topological state.

The emergence of this distinct entanglement phase points to a breakdown of short-range locality in the entanglement Hamiltonian itself.
By reference to fundamental theorems applying to short-range-interacting spin systems, we can infer that the mechanism driving this dissociation results from the emergence of effective long-range interactions in the entanglement Hamiltonian. In the easy-plane regime ($\Delta<1$), these interactions are sufficiently strong to stabilize U(1) symmetry breaking, leading to a unique ground state in the $S_A^z=0$ sector separated by a finite neutral gap at $k=\pi$. This constitutes a direct violation of the standard Lieb--Schultz--Mattis and Mermin--Wagner--Hohenberg--Coleman constraints for short-range systems. 
From a field-theoretic perspective, this breakdown of short-range locality is consistent with the limits of the Bisognano--Wichmann framework~\cite{BW1975,BW1976,Giudici2018,Rottoli2025}: although the low-energy theory at the SU(2)-symmetric point is effectively relativistic, the extensive nature of the bipartition invalidates the geometric assumptions of the theorem. This allows the modular (entanglement) Hamiltonian to develop nonlocal couplings and autonomous criticality, a behavior distinct even from exactly solvable long-range models like the Haldane-Shastry chain which retain conformal symmetry~\cite{Haldane1992}. In contrast, for $\Delta>1$ the interactions become effectively short-ranged, and the LSM scenario is restored via a doubly degenerate ground state. Our results thus provide a concrete benchmark for the recently proposed entanglement LSM-type constraint~\cite{NSR2025}, clarifying the physical conditions under which its predictions hold or fail.
Given that the possibility of long-range interactions is a generic feature of entanglement Hamiltonians, our findings establish a paradigm for understanding entanglement phase transitions across a broad class of quantum many-body systems.

Our findings establish a new frontier of \textit{entanglement-only quantum criticality} as a phase boundary in entanglement space that is not accompanied by any singularity in bulk observables. This behavior is distinct from previously known departures between entanglement and bulk physics: earlier studies revealed criticality only by altering the bipartition of a fixed state~\cite{Hsieh2014}, or identified ES degeneracies as SPT fingerprints~\cite{Pollmann2010}. While recent work has also found differences between entanglement and edge degeneracies in lattice gauge theories~\cite{Pollmann_PRX2025}, our work adds a distinct, experimentally relevant mechanism: a continuous tuning of a microscopic coupling can drive the entanglement Hamiltonian through a genuine phase transition while the bulk remains gapped and non-critical.


Looking forward, the distinct phases of the entanglement Hamiltonian identified here need not remain purely theoretical. Emerging tools for entanglement spectroscopy and tomography in quantum simulators provide promising routes toward experimentally reconstructing entanglement structures~\cite{tomography,Foss-Feig_PRL2022,Quantum}. Furthermore, recent advances in resonant inelastic X-ray scattering (RIXS) have demonstrated experimental access to extensive-cut many-body entanglement through spectroscopic and interferometric techniques~\cite{RIXS,RIXS-PRL,RIXS-Nmaterials}. Together with prospective approaches based on state tomography from neutron scattering data~\cite{Quintanilla2022,Tula2025}, these developments suggest that experimentally probing entanglement-Hamiltonian physics, and ultimately entanglement-only quantum criticality, may become feasible in the future.

\begin{acknowledgments}
\textit{Acknowledgments.}-- We are grateful to Masaki Oshikawa and Frank Pollmann for invaluable discussions.
YCT is grateful for the support of the National Science and Technology Council (NSTC) of Taiwan under grant No.~113-2112-M-A49-015-MY3. GM acknowledges support from the EPSRC under grant no.~EP/P022995/1 and by a University Research Fellowship of the Royal Society under grant no.~URF\textbackslash R\textbackslash180004. We thank the National Center for High-performance Computing (NCHC) of National Applied Research Laboratories (NARLabs) in Taiwan for providing computational and storage resources. We acknowledge use of the ICARUS cluster at the University of Kent, which was partially supported by EPSRC grant No.~EP/S017755/1.
\end{acknowledgments}

\bibliography{ref}

\section*{End Matter}
\subsection{Breakdown of the Lieb--Schultz--Mattis Argument for the Entanglement Hamiltonian}
In this end matter, we revisit the Lieb--Schultz--Mattis (LSM) argument and show that its conclusion fails for the entanglement Hamiltonian due to the breakdown of the locality assumption--not because the theorem is incorrect, but because $H_E$
 violates the short-range interaction condition required for the theorem to apply. 
 
First,
we briefly review the basic setup for the proof of the LSM theorem~\cite{book:Auerbach,LSM1961, LSM:Oshikawa1997,LSM:Oshikawa2000, Kobayashi2019,Yao2021,LSM:Yao2024,Zang2024,Su2025, Ma2024, Liu2025}, which proceeds by defining a \textit{twist operator}. Applying this construction to the case of the entanglement Hamiltonian for subsystem A, we define the twist operator
\begin{equation}
\mathcal{O}_A=\exp\Big(i\frac{2\pi}{L}\sum_{j=1}^LjS_{A,j}^z\Big),
\end{equation}
which rotates each spin incrementally in the $x$-$y$-plane, such that between the first and last site, the spin coordinates are rotated by $2\pi$ about the $z$~axis. The twisted state is constructed as $|\xi_T\rangle=\mathcal{O}_A|\xi_0\rangle$, where $|\xi_0\rangle$ is the entanglement ground state, satisfying $\rho_A|\xi_0\rangle=\omega_0|\xi_0\rangle$, and $\omega_0$ is the largest eigenvalue of the reduced density matrix $\rho_A$. Throughout this study, $|\xi_0\rangle$ lies in the quantum number sector $(S_A^z,k)=(0,k_0)$; where for $L=4m$, $k_0=0$, and for $L=4m+2$, $k_0=\pi$. Since $[\mathcal{O}_A,S_A^z]=0$, the twisted state $|\xi_T\rangle$ also lies in the same magnetization sector with $S_A^z=0$.

Within periodic boundary conditions, the lattice translation $\mathcal{T}$ by one site along the chain acts as $\mathcal{T}\mathbf{S}_{A,j}\mathcal{T}^{-1}=\mathbf{S}_{A,j+1}$, $\mathcal{T}\mathbf{S}_{A,L}\mathcal{T}^{-1}=\mathbf{S}_{A,1}$, and $\mathcal{T}|\xi_0\rangle=e^{ik_0}|\xi_0\rangle$.

In the $S_A^z=0$ sector, which contains the entanglement ground state, the transformation follows:
\begin{align}
\mathcal{T}\mathcal{O}_A\mathcal{T}^{-1}&=\mathcal{O}_A\exp(i2\pi S_1^z)\exp\Big(-i\frac{2\pi}{L}S_A^z\Big)
=\mathcal{O}_Ae^{i\pi}.
\end{align}
Now we can calculate the eigenvalue of $\mathcal{T}$ corresponding to the twisted state $|\xi_T\rangle$.
\begin{align}\label{eq:TOT}
\mathcal{T}|\xi_T\rangle&=\mathcal{T}\mathcal{O}_A|\xi_0\rangle
=(\mathcal{T}\mathcal{O}_A\mathcal{T}^{-1})\mathcal{T}|\xi_0\rangle\notag\\
&=\mathcal{O}_Ae^{i\pi}e^{ik_0}|\xi_0\rangle
=e^{i(k_0+\pi)}\mathcal{O}_A|\xi_0\rangle\notag\\
&=e^{i(k_0+\pi)}|\xi_T\rangle
\end{align}
This shows that the twisted state carries momentum shifted by $\pi$.

The proof of the LSM theorem contains the following two steps:

(i) \textit{Orthogonality}: Prove that the twisted state is orthogonal to the entanglement ground state.
\begin{align}
&\langle\xi_0|\xi_T\rangle=\langle\xi_0|\mathcal{O}_A|\xi_0\rangle
=\langle\xi_0|(\mathcal{T}\mathcal{O}_A\mathcal{T}^{-1})|\xi_0\rangle\notag\\
&=\langle\xi_0|\mathcal{O}_Ae^{i\pi}|\xi_0\rangle=-\langle\xi_0|\xi_T\rangle,
\end{align}
so this scalar product must vanish.
In other words, $|\xi_T\rangle$ serves as a valid variational trial state for the low-lying excitations. By the variational principle, its energy expectation value provides an upper bound on the first excited state energy, i.e. $\xi_1 \le \langle\xi_T|H_E|\xi_T\rangle$.

(ii) \textit{Energy Bound and Breakdown of Locality:}
Under the assumptions of the LSM theorem, namely translation invariance and sufficiently local interactions, the twist construction produces a state that is orthogonal to the ground state while incurring a vanishing energy cost in the thermodynamic limit. This implies that the spectrum must become degenerate (gapless) at momentum $k=\pi$.

Although the entanglement Hamiltonian $H_E$ is often assumed to be local~\cite{NSR2025}, our numerical results (Fig.~\ref{fig3:gap}(b)) demonstrate that this assumption fails in the $\Delta<1$ regime. While the orthogonality condition is satisfied, the neutral gap $\Gamma_n(0)$ remains finite. This indicates that the second step of the LSM argument---the vanishing of the variational energy cost---is violated due to the emergence of effective long-range interactions.

Physically, the variational energy difference is given by
\begin{equation}
\Delta E = \langle \xi_0 | \mathcal{O}_A^\dagger H_E \mathcal{O}_A - H_E | \xi_0 \rangle.
\end{equation}
For interactions decaying as $1/r^\alpha$ in 1D, the energy cost vanishes ($\Delta E \sim 1/L$) only if $\alpha>2$~\cite{Ma2024, Liu2025}. This is because the energy contribution depends on the relative twist angle $\delta\theta_{ij} \sim |i-j|/L$; for $\alpha\le2$, contributions from distant pairs ($|i-j| \sim L$) with large relative twists yield a non-vanishing correction. In the $\Delta<1$ regime, $H_E$ develops interactions that violate this $\alpha > 2$ condition.

\begin{figure*}[t]
\includegraphics[width=0.85\textwidth]{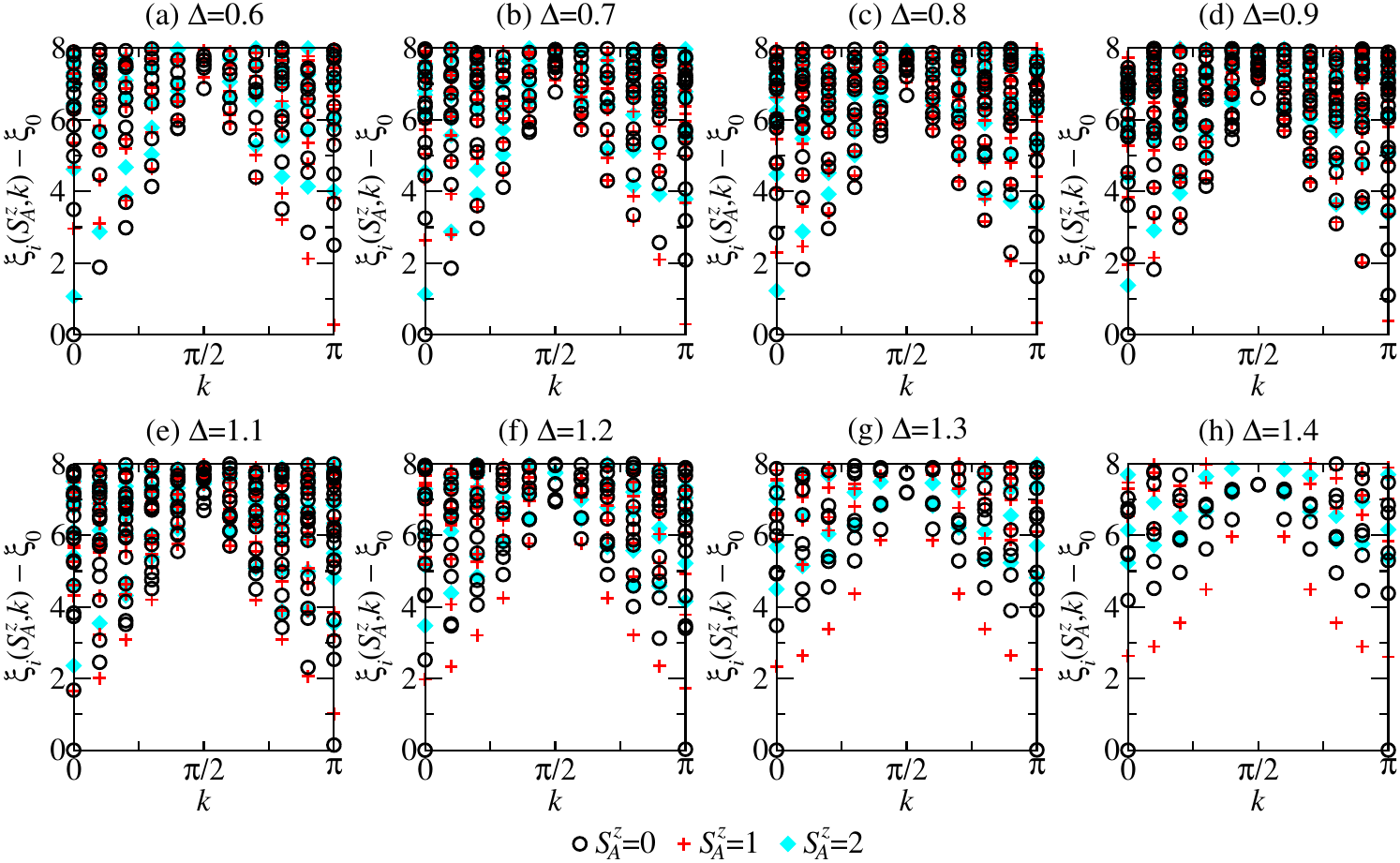} 
\caption{Momentum-resolved entanglement spectra for additional values of $0.5 < \Delta < 1$ and $1 < \Delta < 1.5$, at $J_\mathrm{rung} = -6$, $L = 20$, complementing Fig.\ref{fig:1} in the main text.}\label{fig:S2}
\end{figure*}

Crucially, this threshold $\alpha=2$ is not accidental but fundamental: in 1D systems, long-range interactions with $\alpha \le 2$ are known to invalidate the \textit{entanglement area law}, even for noncritical ground states~\cite{Kuwahara2020}. This rigorous connection highlights that the finite entanglement gap is a generic signature of the breakdown of the area law in the entanglement Hamiltonian. Consequently, tensor-network methods that fundamentally rely on the area law face intrinsic limitations in this regime, underscoring the necessity of unbiased large-scale exact diagonalization. Our results thus provide a crucial benchmark for identifying regimes where standard tensor-network approaches must be supplemented with alternative methods.

Finally, we note that the gapless spectrum reported in Ref.~\cite{NSR2025} corresponds to the SU(2)-symmetric point of the Haldane phase, an entanglement critical point. While the gaplessness at this point mimics the behavior predicted by LSM, it does not imply that the theorem applies universally to $H_E$. In the anisotropic regime ($\Delta<1$), effective long-range interactions invalidate the LSM variational bound, so the finite neutral gap coexists with the presence of an underlying Anderson tower with a parabolic dispersion whose prefactor collapses with system size, highlighting the need for caution when extrapolating to the thermodynamic limit. Rather, this gaplessness is a critical-point phenomenon; away from $\Delta=1$, the locality assumption fails.

\subsection{ES for additional values of $\Delta$}
Figure \ref{fig:S2} supplements Fig.\ref{fig:1} of the main text by showing the momentum-resolved entanglement spectra for intermediate anisotropy values in smaller increments \mbox{$\Delta=0.6,0.7,0.8,0.9,1.1,1.2,1.3,1.4$} at fixed $J_\mathrm{rung}=-6$ and $L=20$. These data fill in the gaps between the representative points $\Delta=0.5,1.0,1.5$ shown in Fig.~\ref{fig:1}, allowing us to track the continuous evolution of the low-lying modes across the entanglement transition.

We note the gradual closing of the secondary gap for the mode at $k_0+\pi$ as $\Delta$ approaches the Heisenberg point $\Delta=1$ from below. On the other side of the phase transition of entanglement, as diagnosed in the main text, note that the two-fold near-degeneracy is already fully developed at $\Delta=1.1$, while the gap continues to open gradually as $\Delta$ increases.

\end{document}